\documentclass[aps,prd,superscriptaddress,nofootinbib,floatfix,preprint]{revtex4-1}

\usepackage[dvipdfmx]{graphicx}
\usepackage{mediabb}
\usepackage{amsmath}
\usepackage{bm}
\usepackage[small,bf]{caption}
\usepackage[subrefformat=parens]{subcaption}
\usepackage{comment}

\newcommand{\beq}{\begin{equation}}
\newcommand{\eeq}{\end{equation}}
\newcommand{\beqa}{\begin{eqnarray}}
\newcommand{\eeqa}{\end{eqnarray}}
\newcommand{\bpr}{\begin{problem}}
\newcommand{\epr}{\end{problem}}
\newcommand{\bcent}{\begin{center}}
\newcommand{\ecent}{\end{center}}
\newcommand{\bfig}{\begin{figure}}
\newcommand{\efig}{\end{figure}}
\newcommand{\bpc}{\begin{picture}}
\newcommand{\epc}{\end{picture}}
\newcommand{\barr}{\begin{array}}
\newcommand{\earr}{\end{array}}
\newcommand{\bitm}{\begin{itemize}}
\newcommand{\eitm}{\end{itemize}}
\newcommand{\bright}{\begin{flushright}}
\newcommand{\eright}{\end{flushright}}
\newcommand{\bminip}{\begin{minipage}}
\newcommand{\eminip}{\end{minipage}}
\newcommand{\btab}{\begin{tabular}}
\newcommand{\etab}{\end{tabular}}

\newcommand{\hiroshima}{Graduate School of Advanced Science and Engineering, Hiroshima University, Kagamiyama, Higashi-Hiroshima 739-8526, Japan}

\newcommand{\icr}{Institute for Chemical Research, Kyoto University Uji, Kyoto 611-0011, Japan}
\newcommand{\tokai}{Research Institute of Science and Technology, Tokai University, 4-1-1 Kitakaname, Hiratsuka, Kanagawa 259-1292, Japan}

\newcommand{\kyoto}{Graduate School of Science, Kyoto University, Sakyouku, Kyoto 606-8502, Japan}

\newcommand{\om}{\omega}

\begin{document}
\title{
Pilot search for axion-like particles by a three-beam stimulated 
resonant photon collider with short pulse lasers
}

\author{Fumiya Ishibashi}\affiliation{\hiroshima}
\author{Takumi Hasada}\affiliation{\hiroshima}
\author{Kensuke Homma\footnote{corresponding author}}\affiliation{\hiroshima}
\author{Yuri Kirita}\affiliation{\hiroshima}
\author{Tsuneto Kanai}\affiliation{\icr}
\author{ShinIchiro Masuno}\affiliation{\icr}
\author{Shigeki Tokita}\affiliation{\icr}\affiliation{\kyoto}
\author{Masaki Hashida}\affiliation{\icr}\affiliation{\tokai}

\date{\today}

\begin{abstract}
Toward the systematic search for axion-like particles in the eV mass range,
we proposed the concept of a stimulated resonant photon collider by 
focusing three short pulse lasers into vacuum.
In order to realize such a collider,
we have performed a proof-of-principle experiment with a set of large incident
angles between three beams to overcome the expected difficulty to ensure the space-time overlap 
between short pulse lasers and also established a method to evaluate the bias on the
polarization states, which is useful for a future variable-incident-angle collision system.
In this paper we present a result from the pilot search with the developed system and the method.
The search result was consistent with null. We thus have set the upper limit
on the minimum ALP-photon coupling down to $1.5 \times 10^{-4}$~GeV${}^{-1}$ at the ALP mass of 1.53 eV
with a confidence level of 95\%.
\end{abstract}

\maketitle

\section{Introduction}
Present space observations consistently estimate that 95\% of the energy density
balance of the Universe is occupied by dark matter and dark energy. 
Among the dark components, axion~\cite{AXION} is one of the most rational candidates for
cold dark matter (CDM), which is supposed to be created via spontaneous breaking of 
the Peccei-Quinn symmetry~\cite{PQ} in order to solve the strong CP problem.
Furthermore, among axion-like particles (ALPs) which set free for the relation 
between mass and coupling unlike the QCD axion, 
the {\it miracle} model~\cite{MIRACLE} 
unifying inflation and dark matter into a single pseudoscalar-type 
ALP predicts the ALP mass and its coupling to photons in a range overlapping
with those of the benchmark QCD axion models in the eV mass range.
Moreover, very recently, a scenario of thermal production of cold "hot dark matter"~\cite{Yin2023} 
and a new kind of axion model from the Grand Unified Theory (GUT) based on 
SU(5) $\times$ U(1)${}_{PQ}$~\cite{TakahashiYin2023} predict ALPs in the eV mass range as well.

In this paper, we thus focus on the following interaction Lagrangian
between a pseudoscalar-type ALP, $\phi_a$, and two photons
\beq\label{Lagrangian}
-{\cal L} = \frac{1}{4}\frac{g}{M}F_{\mu\nu}\tilde{F}^{\mu\nu}\phi_a
\eeq
where
$F^{\mu\nu}=\partial^{\mu}A^{\nu}-\partial^{\nu}A^{\mu}$ is the field strength tensor
and its dual $\tilde{F}^{\mu\nu} \equiv \frac{1}{2}\epsilon^{\mu\nu\alpha\beta}F_{\alpha\beta}$ with
the Levi-Civita symbol $\epsilon^{ijkl}$,
and $g$ is a dimensionless constant while $M$ is an energy at which
a global continuous symmetry is broken.
Specifically, the ALP mass range in ${\cal O}(0.1-1)$~eV at the coupling
$g/M \sim 10^{-11}$~GeV${}^{-1}$ is the intended range of this study.
Therefore, a typical photon energy of laser fields, ${\cal O}(1)$~eV, is
suitable for a photon collider targeting this mass range.

\begin{figure}
\centering
\includegraphics[width=0.8\textwidth]{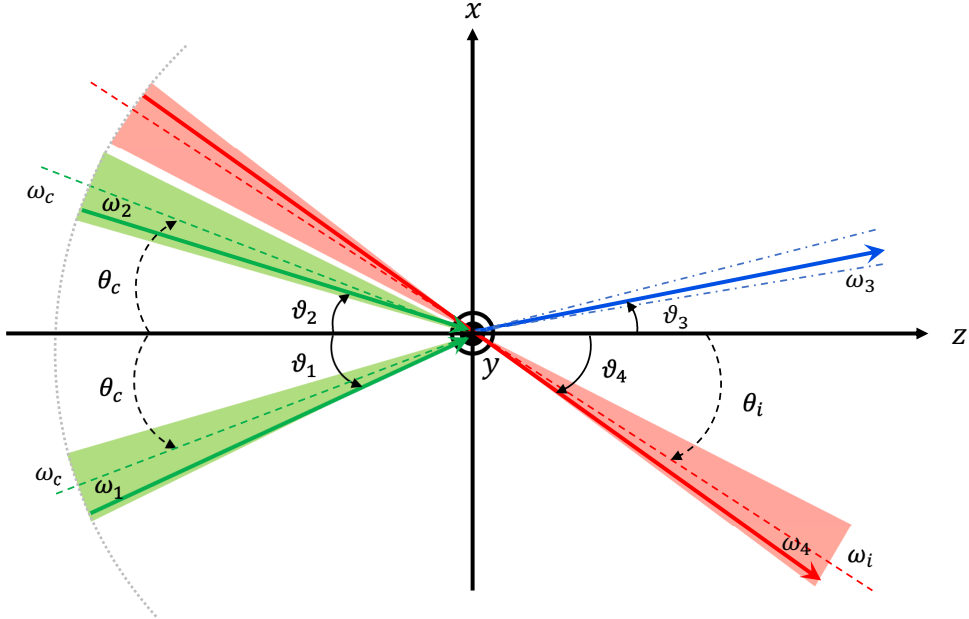}
\caption{
Concept of a three-beam stimulated resonant photon collider (${}^\mathrm{t}$SRPC)
with focused coherent beams.
}
\label{Fig1}
\end{figure}

We have proposed a method to directly produce an ALP resonance state and 
simultaneously stimulate its decay by combining two-color (creation and inducing) 
laser fields and focusing them together with a single lens element in vacuum,
which is defined as stimulated resonant photon collisions (SRPC) in 
a quasi-parallel collision system (QPS)\cite{DEptp}.
In order to satisfy a resonance condition for the direct production of an ALP, 
the range of the center-of-mass system energy, $E_{cms}$, between two photons selected 
from a focused creation laser beam must include the ALP mass, $m_a$.
Thus the condition is simply expressed as
\beq
E_{cms} = 2\omega_c \sin\theta_c = m_a
\eeq
with a common creation laser photon energy $\omega_c$ and 
an angle $2\theta_c$ between the two photons.
Since a typical photon energy in lasers is around eV,
SRPC in QPS has been employed as a way to access sub-eV ALPs with a long focal length
\cite{PTEP2014, PTEP2015, PTEP2020, JHEP, SAPPHIRES00, SAPPHIRES01}.

In order to access a higher mass range above eV,
we have extended the formulation for SRPC with a single focused beam after combining two
lasers in QPS~\cite{UNIVERSE} to SRPC with three separated focused beams 
(${}^\mathrm{t}$SRPC)~\cite{3beam00} as illustrated in Fig.\ref{Fig1}.
We can introduce a symmetric incident angle of $\theta_c$ for the two beam axes of
creation lasers (green), however, two incident photons from the focused two beams
indeed have different incident angles $\vartheta_1$ and $\vartheta_2$ from $\theta_c$
with different energies $\omega_1$ and $\omega_2$ from $\omega_c$, respectively, in general.
The incident angle fluctuations around the beam axes are caused by momentum fluctuations
at around the focal point, while energy uncertainties are caused by nearly Fourier
transform limited short-pulsed lasers. These fluctuations are, in principle,
unavoidable due to the uncertainty principle in momentum-energy space.
Accordingly the exact resonance condition is modified as
\beq
E_{cms} = 2\sqrt{\omega_1\omega_2} \sin\left(\frac{\vartheta_1+\vartheta_2}{2}\right) = m_a.
\eeq
The inducing beam with the central energy $\omega_i$ (red) is simultaneously focused into
the overlapping focal points between the two creation beams and part of the beam 
represented as $\omega_4$ enhances the interaction rate of the stimulated scattering 
resulting in emission of signal photons with the energy $\omega_3$ (blue)
which satisfies energy-momentum conservation.
In order to reflect realistic energy and momentum distributions in the three beams,
numerical calculations are eventually required to evaluate
the stimulated interaction rate~\cite{3beam00}.
Thanks to the broadening of $E_{cms}$ due to these uncertainties, however,
the sensitivity to a target ALP mass also will have a wide resolution around the mass, 
which allows a quick mass scan if we vary $\theta_c$ with a consistent step
with the mass resolution.

On the other hand, synchronization of tightly confined
pulses in space-time is required for ${}^\mathrm{t}$SRPC,
which increases the experimental difficulty.
In a photodetector with electric amplification, the time resolution
is ${\cal O}(10)$~ps at most. 
For the duration of creation laser pulses about 40~fs, such a conventional
detection technique is not applicable for ensuring synchronization of creation laser beams.
Therefore, we consider utilization of nonlinear optical effects in a thin BBO crystal.
Second harmonic generation (SHG) via the 2nd order nonlinear optical effect in BBO
can be used for the synchronization between two creation beams.
As for the three beam synchronization,
the 3rd order nonlinear optical effect, four-wave mixing (FWM), in the same crystal can be used.
For the purpose of synchronization the atomic processes are quite important,
while the atomic FWM becomes the dominant background source 
with respect to FWM in vacuum, that is, generation of 
$\omega_3$ photons via ALP-exchange in ${}^\mathrm{t}$SRPC.
This is because both atomic and ALP-exchange processes require energy-momentum
conservation between four photons and the signal photon energy $\omega_3$ becomes 
kinematically almost identical.

In this paper, we will present a result of the pilot ALP search based ${}^\mathrm{t}$SRPC
in air as a proof-of-principle experiment that
demonstrates the aforementioned method practically works to guarantee 
the space-time synchronization between the three beams
by setting a large collision angle of the creation lasers at $\theta_c=30$~$\deg$
to learn the real technical complications 
toward the continuous mass scanning by systematically varying $\theta_c$ 
in the near future search.
 
In the following sections, we describe the experimental setup and 
the synchronization methods in the pilot ALP search, 
the method for analyzing the acquired data, how to set the exclusion limits,
and, finally, conclude the search result and discuss future plans
toward the continuous ALP mass scanning.


\section{Experimental setup}\label{Sec2} %
\begin{figure}
\centering
\includegraphics[width=1.0\textwidth]{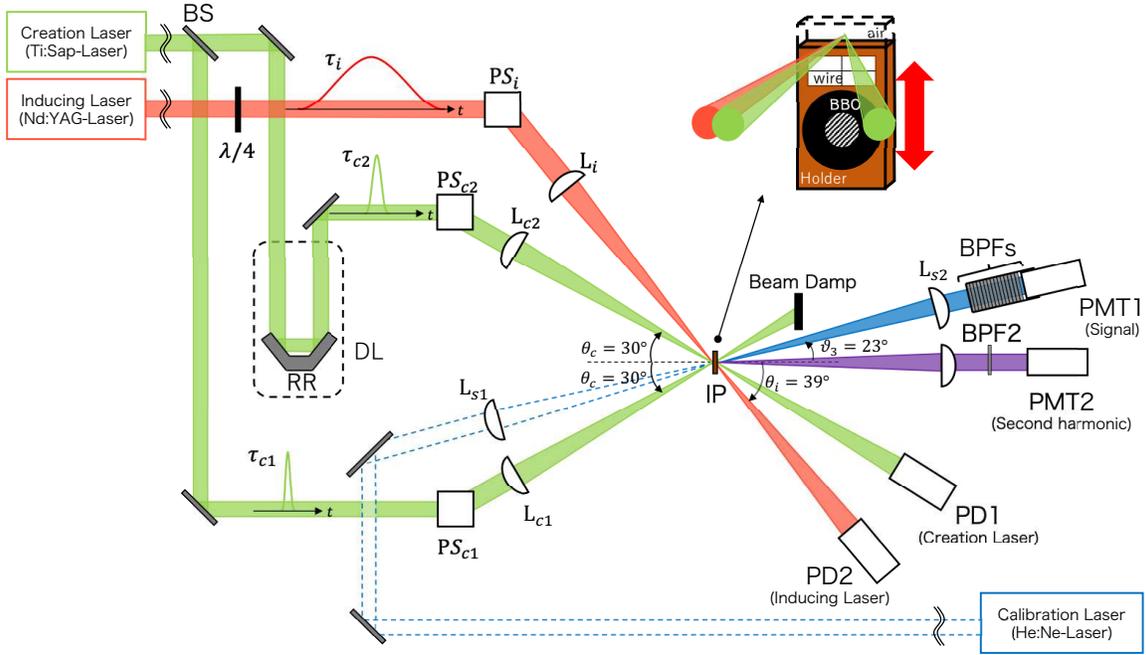}
\caption{Schematic drawing of the search setup.}
\label{Fig2a}
\end{figure}
\begin{figure}
\centering
\includegraphics[width=1.0\textwidth]{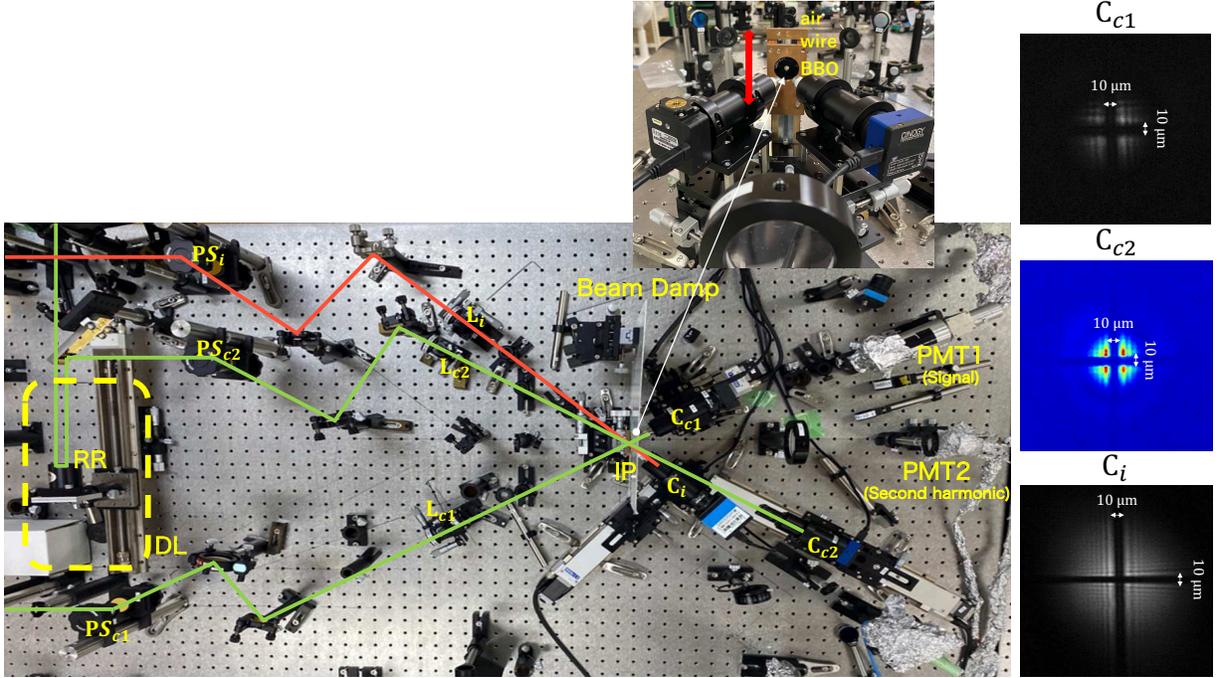}
\caption{Photographs of the search setup (left) and the three focused laser spots (right)
at a common thin cross-wire target.}
\label{Fig2b}
\end{figure}

Figure \ref{Fig2a} and \ref{Fig2b} show a schematic drawing of the searching setup and 
the photographs of the setup with the three focused laser spots 
at a thin cross-wire target, respectively. 
We used a Ti:Sapphire laser (T$^{6}$-system) with $\sim$40~fs duration and a Nd:YAG laser
with 9~ns duration for the creation and inducing fields, respectively. 
Both of them are available in the Institute for Chemical Research in Kyoto University. 
The central wavelengths of these lasers were 808 nm and 1064 nm, respectively. 
Creation laser pulses were injected into a beam splitter (BS) and 
bifurcated to prepare for two creation fields with the guaranteed synchronization.
In this case, one of the creation lasers transmits BS, so the duration of the pulse is slightly elongated. 
Therefore, in principle, there is a finite duration difference in the two pulses ($\tau_{c1}, \tau_{c2}$).
The central wavelength of signal photons is expected to be 651 nm via FWM: 
$\omega_{c1} + \omega_{c2} - \omega_{i}$ with creation photon energies $\omega_{c1}$ and $\omega_{c2}$, respectively,
and inducing photon energy $\omega_{i}$.
In addition to energy conservation, momentum conservation requires the following angle relation:
$\theta_i=39.1^{\circ}$ and the most probable $\vartheta_3=22.7^{\circ}$ for $\theta_c=30.0^{\circ}$
resulting in the resonant mass $m_a =1.53$~eV with respect to the given central photon energies.
Spacetime synchronization at the interaction point (IP) is required 
between two creation pulses branched at BS.
Thus a delay line (DL) equipped with a retroreflector (RR) was constructed on an motorized-stage 
at one of the creation laser paths (upper green line in Fig.\ref{Fig2a}). 
By adjusting the position of RR along DL, the timing for the two pulse incidence at IP can be synchronized. 
In contrast, the inducing laser pulses were electrically triggered by a clock source synchronized with
an upstream oscillator dedicated for the creation laser,
and the injection timing was controlled by a Q-switch based on arrival times to two fast photodiodes 
(PD1, PD2) for one of two creation pulses and for inducing pulses by looking at an oscilloscope.

Individual beams were focused into IP via periscopes (PS$_{c1}$, PS$_{c2}$, PS$_{i}$)
at $30^\circ$ for the creation lasers and $39^\circ$ for the inducing laser 
as shown in Fig.\ref{Fig2a}. These incident angles and signal outgoing angle were determined 
so that the central signal wavelength from FWM becomes 651~nm via energy-momentum conservation.

Typically a mirror is designed to maximize reflectivity at an angle of incidence (AOI) of $45^\circ$
and thus a reflection angle of $45^\circ$ which can maintain linearly polarized states 
with respect to linearly polarized incident beams.
A periscope (PS) consists of a pair of mirrors aligned vertically with AOI of $45^\circ$ 
while it can emit a beam in any directions by changing the optical axis (beam height).
Thus, in the near future, we will be able to scan collision angles between the two creation beams by 
the introduction of PS. However, if we use PS to rotate emission directions 
at arbitrarily large angles, polarization states of beams will become elliptic in general. 
Furthermore, one of the creation laser paths contains RR and it can also be a source of
changing elliptical polarization states.
Therefore, it is necessary to introduce complex Jones vectors for representing 
the polarization vectors with two independent angle parameters.
The two angles representing the polarization state of the two creation lasers
were determined by measuring Stokes parameters as explained in Appendix.
On the other hand, the inducing laser was set to circular polarization (left-handed) 
using a $\lambda/4$ plate. This is because the theoretical interface is prepared
for generally polarized states for the creation lasers and circularly polarized
states for the inducing laser~\cite{UNIVERSE} in order to avoid complication on the
numerical calculations to estimate inducible momentum ranges in the final state~\cite{JHEP,UNIVERSE}.

The two creation lasers and the inducing laser were focused at IP with 
lenses L$_{c1}$, L$_{c2}$ and L$_{i}$, respectively,
with a common focal length of $f = 300$~mm as shown in Fig.\ref{Fig2a}.
IP was equipped with a special holder vertically consisting of 
a cross-wire of $10~ \mu$m thickness, a BBO crystal which is a nonlinear crystal of 
$50$~$\mu$m thickness, and a no target state (air) 
as shown in the insets of Fig.\ref{Fig2a} and \ref{Fig2b}.
By attaching this special holder to the z-axis stage, 
cross-wire (spatial overlap), BBO (time synchronization), and 
no target state (search experiment) can be switched independently of the other optical elements.
The camera systems (C${}_{c1}$, C${}_{c2}$, C${}_{i}$) and photodetectors were located downstream from IP. 
Since individual camera systems are installed on motorized-stages, 
they can be moved to appropriate positions for checking the spatial overlap of 
the three beam spots, the time synchronization between the two or three laser pulses
and performing searches, depending on the purposes.
The spatial overlap was ensured by aligning the center of individual laser spots to 
the crossed point of the two thin wires as shown in the three pictures in Fig.\ref{Fig2b}.
The beam waist for the inducing laser was enlarged compared to those of the creation lasers 
so that the creation laser spots could be stably included in the volume of the inducing field.

After ensuring the spatial overlap between the three beams at IP, 
time synchronization was first performed with the two creation lasers. 
The duration of the creation laser pulses was $\sim$~40~fs. It is impossible to ensure synchronization 
using a conventional photodetector due to the limited time resolution at most $\sim 10$~ps. 
Therefore, space-time synchronization was confirmed by observing second harmonic generation 
from the BBO crystal, which is known as a fast nonlinear optical effect 
with $\sim$~fs resolution when two high-intensity pulses spatiotemporally overlap.
DL was actually adjusted by measuring the number of second harmonic photons as a function of RR position.
In addition to the two creation pulse overlap, when the inducing laser spatiotemporally overlaps with
the creation pulses, FWM in BBO may also be produced. 
Second harmonic from the two creation pulses and FWM from the three pulse overlap emerge 
at different angles. We note that FWM must conserve energy-momentum, while second harmonic
conserves energy but not necessarily momentum because translation symmetry is 
broken in the BBO crystal.

Second harmonic was detected using a photomultiplier tube (PMT2) by selecting second
harmonic of the creation laser wavelength, 404~nm, by a band-pass filter (BPF2). 
Fifteen band-pass filters (BPFs) were placed in front of the PMT for FWM detection (PMT1) 
in order to mainly remove residual beam photons. 
The BPFs were installed in multiple layers of three types of BPFs so that 
they eliminate wavelengths of the creation laser and second harmonic of the creation laser, 
the inducing laser and its second harmonic.
In this way, PMT1 can detect photons only in the proper energy band consistent with FWM. 
Second harmonic and FWM photons from BBO ensured the space-time synchronization of the three lasers.
Since the duration of the inducing laser was 9~ns,
the time resolutions of typical photo-device, $\sim 40$~ps was sufficient to adjust arrival time difference 
between second harmonic and FWM photons, both of which were measured with photomultipliers 
with the same time resolution of 0.75~ns.
After the space-time synchronization between the three beams was ensured, 
the vertical position of the holder was set at the no-target position
and we conducted the search experiment.

\section{Space-time synchronization}
As shown in the pictures of the three beam spots in Fig.\ref{Fig2b} (right),
the centers of the three beams focal spots were adjusted at the crossed point of the crossed wires
of 10~$\mu$m thickness. This guarantees the spatial overlap between the three beams.

\begin{figure}
\centering
\includegraphics[width=0.8\textwidth]{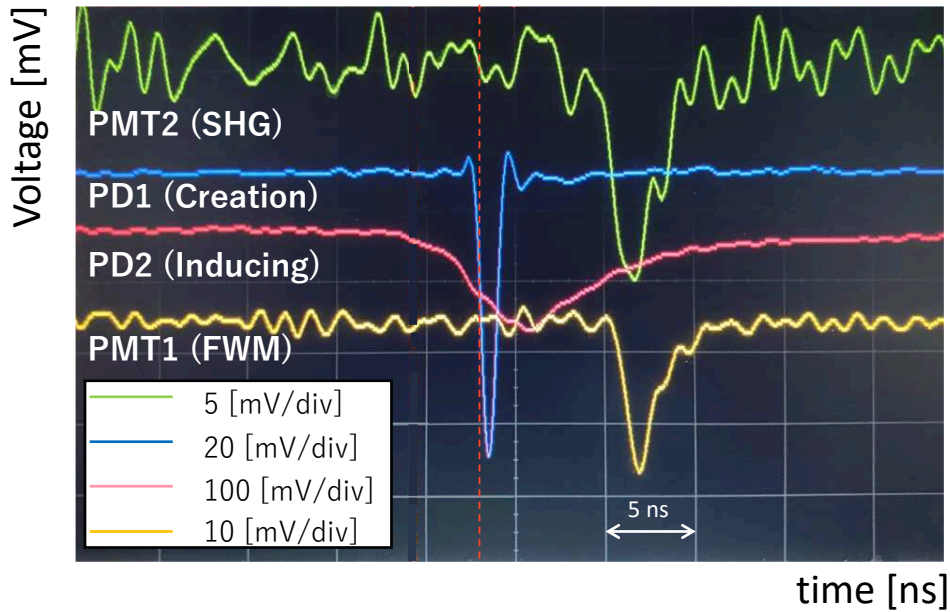}
\caption{
Photograph of oscilloscope waveforms from the four photo-detectors in Fig.\ref{Fig2a}.
Four-wave mixing (FWM) photons were clearly observed
when a thin BBO crystal was positioned at IP.
}
\label{Fig4}
\end{figure}

Figure \ref{Fig4} shows a picture of oscilloscope waveforms when space-time synchronization between
three beams were satisfied, where photodiode signals PD1 (creation laser), PD2 (inducing laser)
and signals from photomultipliers PMT2 (second harmonic generation from BBO),
PMT1(four-wave mixing from BBO) in Fig.\ref{Fig2a} are simultaneously displayed.
When the BBO crystal was inserted to the position of IP, we clearly confirmed
the time synchronization between the three beams.

\begin{figure}
\centering
\includegraphics[width=0.8\textwidth]{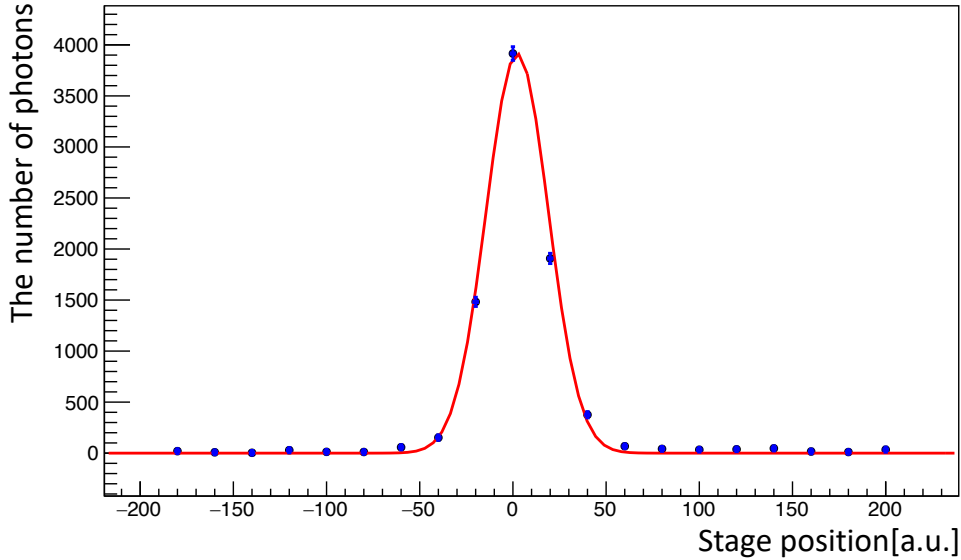}
\caption{
Observed number of FWM photons as a function of stage position in the delay line
for a fine timing tuning between two creation laser pulses 
when a thin BBO crystal was positioned at IP.
}
\label{Fig5}
\end{figure}

For a fine timing tuning between the two creation short pulses, 
we took a look at the number of FWM photons detected by PMT1
as a function of stage position in the delay line (DL) in Fig.\ref{Fig2a}.
Figure \ref{Fig5} shows the clear peak structure at the best synchronization point.


\section{Data analysis}

        \begin{figure}
        \centering
        \includegraphics[width=0.8\textwidth]{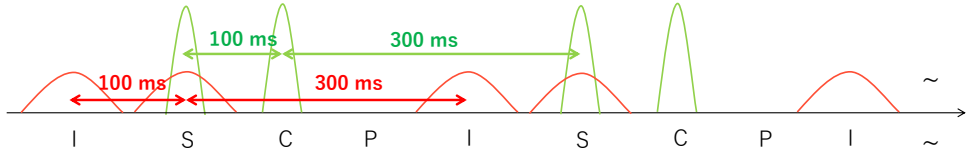}
        \caption{
Four patterns of the beam combination between the two laser pulses
where the green and red pulses are respectively creation and inducing laser pulses.
The classifications are:
S for two laser pulses, C for only the creation laser pulses, I for only the inducing laser pulses, 
and P for pedestals without laser pulses.
}
        \label{sicp_power}
        \end{figure}


PMT1 detects photons from various background sources in addition to signal photons via FWM.
The number of photons detected by PMT1 contains photons or photon-like events
in the following four categories:
the number of signal photons, $n_{sig}$, originatinga from the combination
between the creation and inducing laser pulses,
the number of background photons, $n_c$, originating from only the creation laser pulses, 
the number of background photons, $n_i$, originating from only the inducing laser pulses, 
and
the number of noise photons, $n_{p}$, when no beam exists, that is, pedestal.
In order to extract the number of signal photons, 
the number of photons in the above three background categories must be subtracted.
Therefore, in the search experiment,
both the creation and inducing laser pulses were injected at different irregular intervals of 5 Hz 
as illustrated in Fig.\ref{sicp_power} in order to successively form the four patterns.
The number of measured photons in each pattern is expressed as Eq.\eqref{sicp_pattern}.
The number of photons detected in P-pattern, $N_P$, is the pedestal component from 
environmental noises including thermal noise from PMT1.
The number of photons detected in C- and I-pattern, $N_C$ and $N_I$, respectively, 
include the number of photons originating from individual laser focus 
such as plasma creation on top of the pedestals.
The number of photons detected in S-pattern, $N_S$, includes the number of signal photons 
on top of all the other background sources.

        \begin{eqnarray}
        \begin{split}
        \label{sicp_pattern}
        N_{S} &= n_{sig} + n_{c} + n_{i} + n_{p}\\
        N_{C} &= n_{c} + n_{p}\\
        N_{I} &= n_{i} + n_{p}\\
        N_{P} &= n_{p}
        \end{split}
        \end{eqnarray}

These four patterns were substituted into Eq.\eqref{signal_logic}, 
in order to extract the observed number of FWM photons

        \begin{equation}
        \label{signal_logic}
         n_{obs} = N_{S} - (N_{C}-N_P) - (N_{I}-N_P) - N_{P}.
        \end{equation}

\label {time_window}

        \begin{figure}
        \centering
        \includegraphics[width=0.5\textwidth]{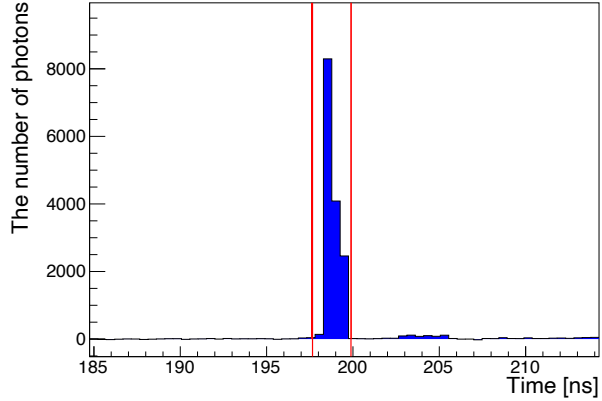}
        \caption{
Arrival time distribution of FWM photons via the atomic process 
when a BBO crystal was placed at IP and space-time synchronization was ensured by PMT1. 
The red lines thus provide the expected time window for FWM photons via
ALP-exchange to arrive. 
}
        \label{bbo_time}
        \end{figure}

In the search, the two photodiodes (PD1, PD2) were placed downstream of the interaction point (IP)
in Fig.\ref{Fig2a}.
Four patterns, S, I, C and P, were defined based on analog waveforms obtained from 
PD1 and PD2 assigned for the creation and inducing lasers, respectively.
The number of photons was reconstructed from the voltage-time relation of analog signals
from PMT1 with a waveform digitizer and applying a peak-finding algorithm to simultaneously determine
the number of photons and their arrival times from falling edges of amplitudes of waveforms.
The details of these instruments and the peak analysis method are described 
in~\cite{SAPPHIRES00}.

In advance of the search, the expected arrival time of FWM photons in vacuum
was determined by the arrival timing of FWM photons in BBO, 
which ensures space-time synchronization between focused three laser pulses. 
Figure \ref{bbo_time} shows the arrival time distribution of FWM photons from BBO, 
where 1000 shots in S-pattern without background subtraction from the other patterns is shown.
In the following analysis, $n_{obs}$ always implies the number of observed FWM photons 
by integrating photon-like charges in PMT1 within the arrival time window
of 2.5 ns which is indicated by the two vertical lines in Fig.\ref{bbo_time}. 

\section{Search result}
\label {signal_analysis}
        \begin{figure}[t]
        \centering
        \includegraphics[width=0.8\textwidth]{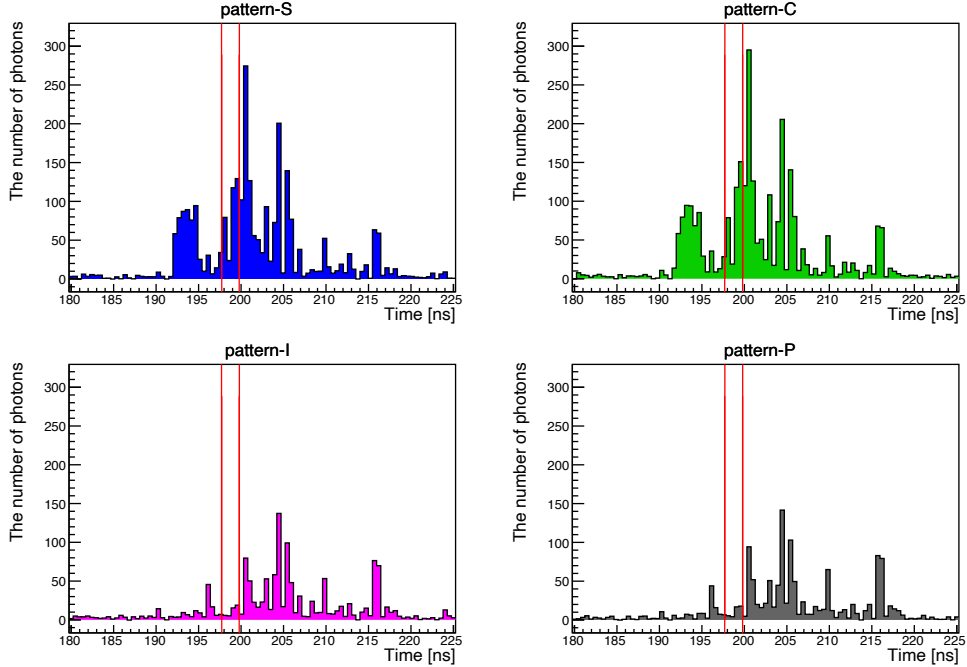}
        \caption{
Arrival time distributions of photons with no target state (air) at IP. 
The histograms in the upper left, upper right, lower left and lower right correspond to 
S, C, I and P patterns of beam combinations, respectively. 
The interval between the two red lines in S-pattern indicates the expected time windows 
for FWM photons via ALP-exchange to arrive. 
}
        \label{fwm_pattern}
        \end{figure}

\begin{figure}
        \centering
        \includegraphics[width=0.8\textwidth]{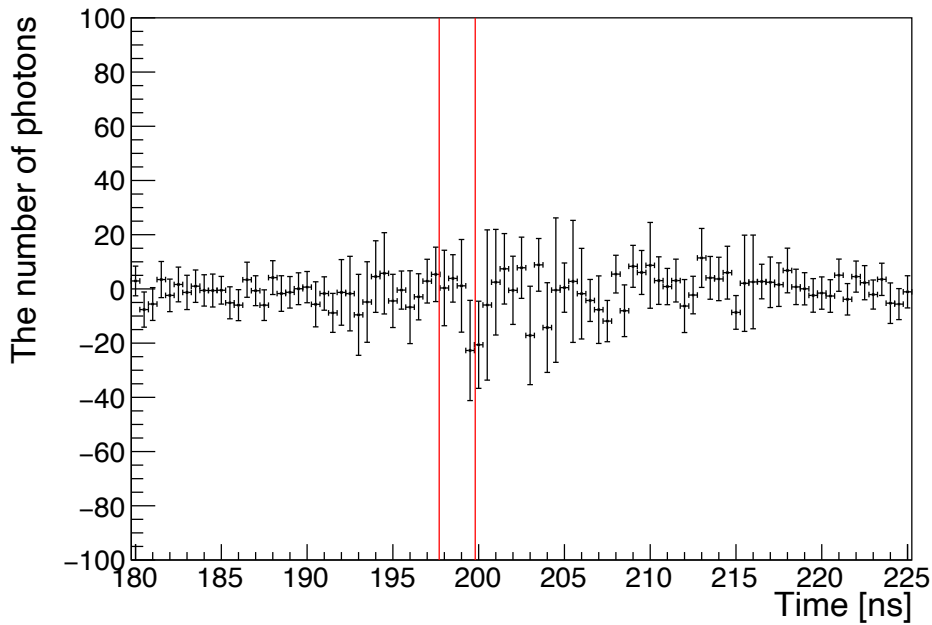}
        \caption{
Arrival time distribution of reconstructed photons after subtraction
between the four patterns based on Eq.\eqref{signal_logic} over the entire time range in
Fig.\ref{fwm_pattern}.
The interval between the two red lines indicates the expected time windows 
for FWM photons via ALP-exchange to arrive. 
}
        \label{null_result}
        \end{figure}

Figure \ref{fwm_pattern} shows arrival time distributions of photons 
in individual patterns. The histograms in S, C, I and P patterns are shown 
in the upper left (blue), upper right (green), lower left (pink), and lower right (gray), 
respectively. The expected arrival time windows were indicated by the two vertical lines. 
The total number of laser shots was 48,000 in the four patterns and thus the valid statistics
in S-pattern was 12,000 shots.
Figure \ref{null_result} shows arrival time distributions after subtruction
with Eq.\eqref{signal_logic}.
The interval between the two vertical lines represents the expected arrival time window 
of FWM photons. Thus the number of FWM photons was evaluated by summing
charges in PMT1 within this window and dividing the sum by single-photon equivalent charge.
As a result, the observed number of FWM photons, $n_{obs}$, was null within the error size
as follows
        \begin{equation}
        \label{signal_photon}
         n_{obs} = -17.4 \pm 28.4 \rm({stat.)} \pm 9.8 \rm{(syst.\mathrm{I})} \pm 5.4 
\rm{(syst.\mathrm{I}\hspace{-1.2pt}\mathrm{I})} +22.4 -15.2 \rm{(syst.\mathrm{I}\hspace{-1.2pt}\mathrm{I}
\hspace{-1.2pt}\mathrm{I})}.
        \end{equation}
The first systematic error (syst.$\mathrm{I}$) was estimated by calculating the root-mean-square 
of the number of photon-like noise excluding the expected arrival time window of FWM photons.
This corresponds to the baseline uncertainty of the PMT1 connected to the waveform digitizer
in the real noise environment.
The second systematic error (syst.$\mathrm{I}\hspace{-1.2pt}\mathrm{I}$) was obtained
by changing the default internal threshold $-1.3$~mV in the peak finder from $-1.2$ to $-1.4$~mV 
with the assumption of  the uniform distribution.
The detail of the peak finding method is explained in \cite{PTEP2020, SAPPHIRES00}.
The third systematic error (syst.$\mathrm{I}\hspace{-1.2pt}\mathrm{I}\hspace{-1.2pt}\mathrm{I}$) was 
evaluated by changing the expected arrival time window size for FWM photons from 1.5 ns to 3.5 ns
with respect to the most likely arrival time window of 2.5 ns.

\section{Exclusion region in the coupling-mass relation for ALP-exchange}
Since we have obtained the null result in Section~\ref{signal_analysis}, we set an exclusion region 
in the coupling-mass relation for the ALP exchange based on the formulation for 
the signal photon yield given in Ref.\cite{3beam00} and the measured total error size as follows.
The signal photons yield in stimulated resonant scattering per pulse collision, $\mathcal{Y}_{c+i}$,
is expressed as~\cite{3beam00}
        \begin{equation}
        \label{yield}
        \mathcal{Y}_{c+i} \equiv N_{1} N_{2} N_{4} \mathcal{D}_{three} \left[s / L^3\right] \bar{\Sigma}_I \left[L^3 / s\right],
        \end{equation}
where $N_{1}(=N_{c1}), N_{2}(=N_{c2})$ and $N_{4}(=N_{i})$ are 
the average numbers of photons contained individual lasers, 
respectively, $\mathcal{D}_{three}$ is a factor representing the space-time overlap of 
focused three beams at the interaction point~\cite{3beam00}
and $\bar{\Sigma}_I$ is the volume-wise interaction rate~\cite{JHEP,UNIVERSE}.
Individual units are given in [~~] with length $L$ and time $s$.

Based on the set of experimental parameters $P$ summarized in Tab.\ref{Tab1},
the observed number of FWM photons via an ALP exchange with the mass $m_{a}$ and the coupling $g /M$ 
to two photons is expressed as
        \begin{equation}
        \label{obtain_photon}
        n_{obs} = \mathcal{Y}_{c+i} \left(m_{a}, g / M; P\right) N_{shot} \epsilon,
        \end{equation}
where $N_{shot}$ is the number of shots in S-pattern and $\epsilon$ is 
the overall detection efficiency.
A coupling constant $g /{M}$ can be evaluated by solving Eq.\eqref{obtain_photon} 
for an ALP mass $m_{a}$ and a given observed number of photons $n_{obs}$.

When counting photon-like peaks by the peak finding algorithm in waveforms, 
fluctuations of the baseline may produce both positive and negative 
amplitudes resulting in negative numbers of photon-like peaks as well as positive ones.
Thus, even if the mean value is zero, we assume a Gaussian distribution as the most
natural null hypothesis. The confidence level for this null hypothesis is defined as 
        \begin{equation}
        \label{confidence_level}
        1 - \alpha = \frac{1}{\sqrt{2 \pi} \sigma} \int_{\mu-\delta}^{\mu+\delta} e^{-(x-\mu)^2 /\left(2 \sigma^2\right)} d x=\operatorname{erf}\left(\frac{\delta}{\sqrt{2} \sigma}\right),
        \end{equation}
where $\mu$ is the expected value of $x$ according to the hypothesis and $\sigma$ 
is the standard deviation. 
In the search, the expected value $x$ corresponds to the number of FWM photons $n_{obs}$
and $\sigma$ is one standard deviation $\delta n_{obs}$.
Based on \eqref{signal_photon} which indicates $\mu=0$ because of the null result,
we determined the acceptance-uncorrected uncertainty $\delta n_{obs}$ 
around $n_{obs}=0$ as the root mean square of all the error components as follows
        \begin{equation}
        \label{Delta}
        \delta n_{obs} = \sqrt{28.4^2 + 9.8^2 + 5.4^2 + 22.4^2} \simeq 37.9
        \end{equation}
where the larger error on the positive side (+22.4) was used for 
syst.$\mathrm{I}\hspace{-1.2pt}\mathrm{I}\hspace{-1.2pt}\mathrm{I}$ in \eqref{signal_photon}.
In order to set a 95\% confidence level, $2\alpha = 0.05$ with $\delta = 2.24$ was used 
to obtain the one-sided upper limit by excluding $x + \delta$~\cite{PDGstatistics}.
The upper limit in the relation $m_{a}$ vs. $g / M$ was estimated by numerically solving 
Eq.\eqref{N_obs} with $\delta n_{obs}$ in \eqref{Delta} 
for the set of experimental parameters $P$ in Tab.\ref{Tab1}
        \begin{equation}
        \label{N_obs}  
        2.24 \delta n_{obs} = \mathcal{Y}_{c+i}\left(m_{a}, g / M; P\right) N_{shot} \epsilon,
        \end{equation}
where $N_{shot}=12,000$ and
the overall efficiency $\epsilon \equiv \epsilon_{opt}\epsilon_{det}$ with
the optical path acceptance from IP to PMT1, $\epsilon_{opt}$, and 
the single photon detection efficiency of PMT1, $\epsilon_{det}$,
were substituted.

\begin{figure}
        \centering
        \includegraphics[width=1.0\textwidth]{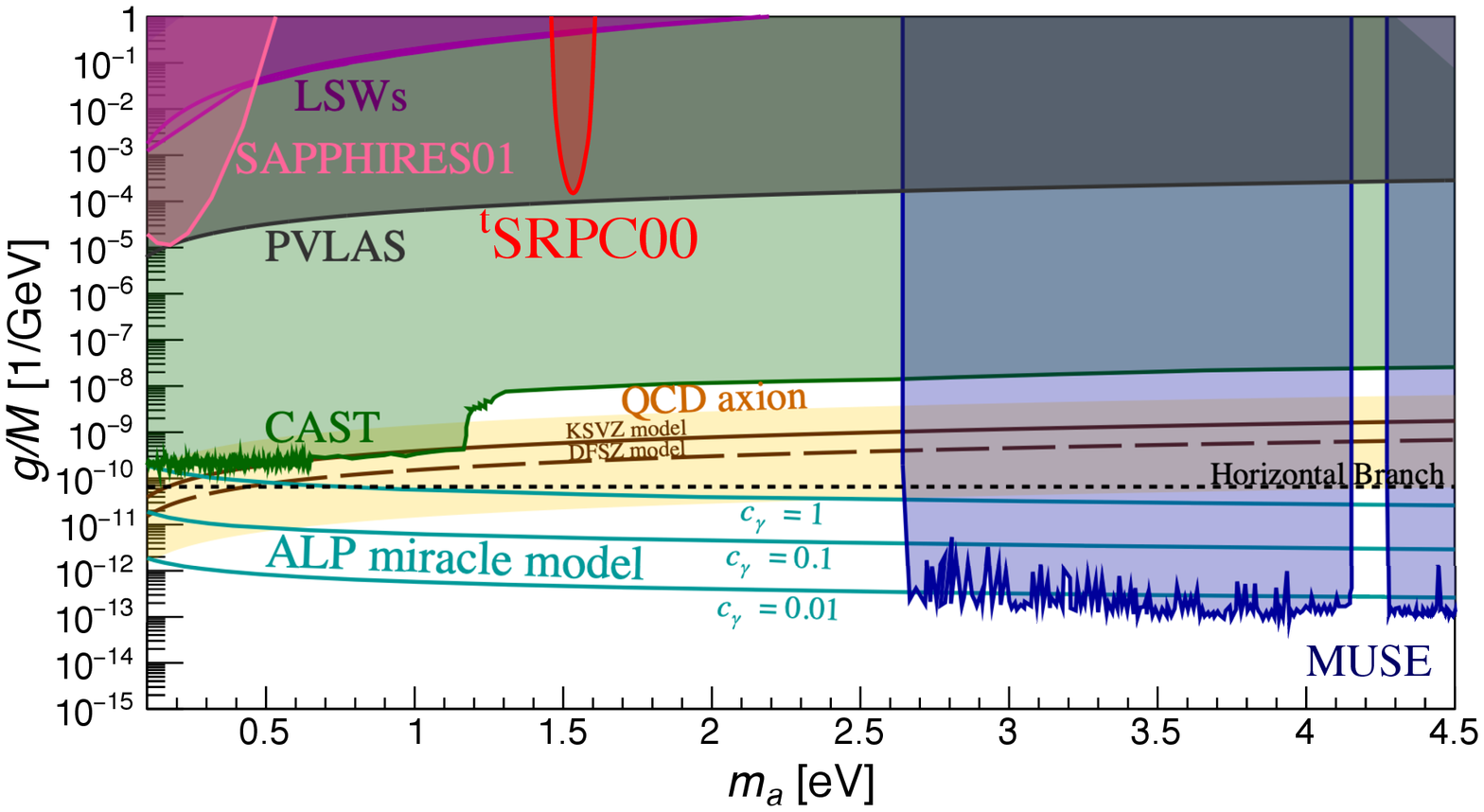}
        \caption{Upper limit in the parameter space of the coupling-mass relation (region enclosed by the red solid curve) evaluated at a 95\% confidence level for the pseudoscalar field exchange
achieved by the three-beam stimulated resonant photon collider (${}^\mathrm{t}$SRPC00).
The magenta area indicates the excluded range based on SRPC in 
quasi-parallel collision geometry (SAPPHIRES01)~\cite{SAPPHIRES01}.
The purple areas are excluded regions by the Light-Shining-through a Wall (LSW) experiments 
(ALPS~\cite{ALPS} and OSQAR~\cite{OSQAR}).
The gray area shows the excluded region by the vacuum magnetic birefringence (VMB) 
experiment (PVLAS~\cite{PVLAS}).
The horizontal dotted line indicates the upper limit from the Horizontal Branch observation \cite{HB}.
The blue areas indicate the excluded regions from the optical MUSE-faint survey \cite{MUSE}. 
The green area is the excluded region by the helioscope experiment CAST \cite{CAST}. 
The yellow band and the upper solid brown line are the predictions of QCD axion 
by the KSVZ model~\cite{KSVZ} with $0.07 < \left|E/N - 1.95\right| < 7$ and $E/N = 0$, respectively. 
The bottom dashed brown line is the prediction from the DFSZ model~\cite{DFSZ} with $E/N = 8/3$. 
The cyan lines are the predictions from the ALP {\it miracle} model~\cite{MIRACLE} 
with the intrinsic parameter values $c_{\gamma}=1, 0.1, 0.01$, respectively.}

        \label{3beam01_ps}
        \end{figure}

Figure \ref{3beam01_ps} shows
the upper limit in the coupling-mass relation from this search, 
the three-beam stimulated resonant photon collider (${}^\mathrm{t}$SRPC00) 
enclosed by the red solid curve. The limit was set 
at a 95\% confidence level by assuming only pseudoscalar-type ALP exchanges.
The most sensitive ALP mass in this search is expected to be $m_{a} = 1.53$~eV,
because the creation lasers have a fixed collision angle of $30^\circ$.
In reality, however, the sensitivity is not limited to $m_{a} = 1.53$~eV
because of energy and momentum uncertainties of focused short pulse lasers.
These uncertainties are exactly taken into account in the numerical calculation
based on Eq.\eqref{yield}~\cite{3beam00}.
The magenta area indicates the excluded range based on SRPC in 
quasi-parallel collision geometry (SAPPHIRES01)~\cite{SAPPHIRES01}.
The purple areas are excluded regions by the Light-Shining-through a Wall (LSW) experiments 
(ALPS~\cite{ALPS} and OSQAR~\cite{OSQAR}).
The gray area shows the excluded region by the vacuum magnetic birefringence (VMB) 
experiment (PVLAS~\cite{PVLAS}).
The horizontal dotted line indicates the upper limit from the Horizontal Branch observation \cite{HB}.
The blue areas indicate exclusion regions from the optical MUSE-faint survey~\cite{MUSE}.
The green area is the excluded region by the helioscope experiment CAST \cite{CAST}.
We also put predictions from the benchmark QCD axion models.
The yellow band and the upper solid brown line are the predictions from
the KSVZ model \cite{KSVZ} with $0.07 < \left|E/N - 1.95\right| < 7$ and
$E/N = 0$, respectively, while the bottom dashed brown line is the prediction 
from the DFSZ model \cite{DFSZ} with $E/N = 8/3$. 
The cyan lines are the predictions from the ALP {\it miracle} model \cite{MIRACLE} 
with the intrinsic parameters $c_{\gamma}=1, 0.1, 0.01$.

        \begin{table}
        \caption{Experimental parameters used to obtain the upper limit.}
        \begin{center}
        \scalebox{1.0}[0.88] {
        \begin{tabular}{lr}  \\ \hline
        Parameter & Value \\ \hline
        Central wavelength of creation laser, $\lambda_{c1}$   & 808 nm\\
        Relative linewidth of creation laser, $\delta\omega_{c1}/<\omega_{c1}>$ &  $1.7\times 10^{-2}$\\
        Duration time of creation laser, $\tau_{c1}$ & (38.8 $\pm$ 1.4) fs (FWHM)\\
        Measured creation laser energy per $\tau_{c1}$, $E_{c1}$ & (1.21 $\pm$ 0.13) $\mu$J \\
        Creation energy fraction within 3~$\sigma_{xy}$ focal spot, $f_{c1}$ & 0.82\\
        Effective creation energy per $\tau_{c1}$ within 3~$\sigma_{xy}$ focal spot & $E_{c1} f_{c1} = 1.0$~$\mu$J\\
        Effective number of creation photons, $N_{c1}$ & $4.0 \times 10^{12}$ photons\\
        Beam diameter of creation laser beam, $d_{c1}$ & (5.0 $\pm$ 0.5)~mm\\
        *Polarization (see Appendix) & $\epsilon_{c1} = 0.41$ rad, $\theta_{c1} = 0.30$ rad \\ \hline
        Central wavelength of creation laser, $\lambda_{c2}$   & 808 nm\\
        Relative linewidth of creation laser, $\delta\omega_{c2}/<\omega_{c2}>$ &  $1.7\times 10^{-2}$\\
        Duration time of creation laser, $\tau_{c2}$ & (39.2 $\pm$ 1.7) fs (FWHM)\\
        Measured creation laser energy per $\tau_{c2}$, $E_{c2}$ & (1.52 $\pm$ 0.14) $\mu$J \\
        Creation energy fraction within 3~$\sigma_{xy}$ focal spot, $f_{c2}$ & 0.85\\
        Effective creation energy per $\tau_{c2}$ within 3~$\sigma_{xy}$ focal spot & $E_{c2} f_{c2} = 1.3$~$\mu$J\\
        Effective number of creation photons, $N_{c2}$ & $5.2 \times 10^{12}$ photons\\
        Beam diameter of creation laser beam, $d_{c2}$ & (5.0 $\pm$ 0.5)~mm\\
        *Polarization (see Appendix) & $\epsilon_{c2} = 0.91$ rad, $\theta_{c2} = - 0.31$ rad \\ \hline 
        Central wavelength of inducing laser, $\lambda_i$   & 1064~nm\\
        Relative linewidth of inducing laser, $\delta\omega_{i}/<\omega_{i}>$ &  $1.0\times 10^{-4}$\\
        Duration time of inducing laser beam, $\tau_{ibeam}$ & 9~ns (two standard deviation)\\
        Measured inducing laser energy per $\tau_{ibeam}$, $E_{i}$ & $(1.58 \pm 0.05)$~$\mu$J \\
        Linewidth-based duration time of inducing laser, $\tau_i/2$ & $\hbar/(2\delta\om_{i})=2.8$~ps\\
        Inducing energy fraction within 3~$\sigma_{xy}$ focal spot, $f_i$ & 0.88\\
        Effective inducing energy per $\tau_i$ within 3~$\sigma_{xy}$ focal spot & $E_{i} (\tau_i/\tau_{ibeam}) f_i = 0.87$~nJ\\
        Effective number of inducing photons, $N_i$ & $4.7 \times 10^{9}$ photons\\
        Beam diameter of inducing laser beam, $d_{i}$ & $(3.0 \pm 0.5)$~mm\\
        Polarization & circular (left-handed state) \\ \hline
        Common focal length of lens, $f$ & 300.0~mm\\
        Single-photon detection efficiency, $\epsilon_{det}$ & 1.4\% \\
        Efficiency of optical path from IP to PMT, $\epsilon_{opt}$ & 53\% \\ \hline
        Total number of shots in trigger pattern S, $N_{shot}$   &  12000~shots\\
        $\delta{n}_{obs}$ & 37.9\\
        \hline
        \end{tabular}
        } 
        \end{center}
        \label{Tab1}
        \end{table}

\section{Conclusion and Future prospect}
We presented a result of the pilot ALP search by 
a three-beam stimulated resonant photon collider (${}^\mathrm{t}$SRPC) with focused short pulse lasers
in air as a proof-of-principle experiment. 
We demonstrated that the space-time synchronization between a pair of short creation
laser pulses with a large incident angle of 30~deg
and a relatively long-duration inducing laser pulse can be ensured 
by atomic four-wave mixing with a thin BBO crystal positioned at the interaction point.
The search result was consistent with null and we could successfully obtained 
an exclusion region in the minimum coupling $g/M = 1.5 \times 10^{-4}$~GeV${}^{-1}$
at $m = 1.53$~eV based on the formulation dedicated for ${}^\mathrm{t}$SRPC~\cite{3beam00}.
We found the solutions to technical complications to handle 
three focused short-pulsed beams and the impact on the physics, in particular,
on the polarization states of creation beams by the introduction of periscopes, which is
an important optical element to realize variable incident angles at a ${}^\mathrm{t}$SRPC. 

\begin{figure}
\centering
\includegraphics[width=0.8\textwidth]{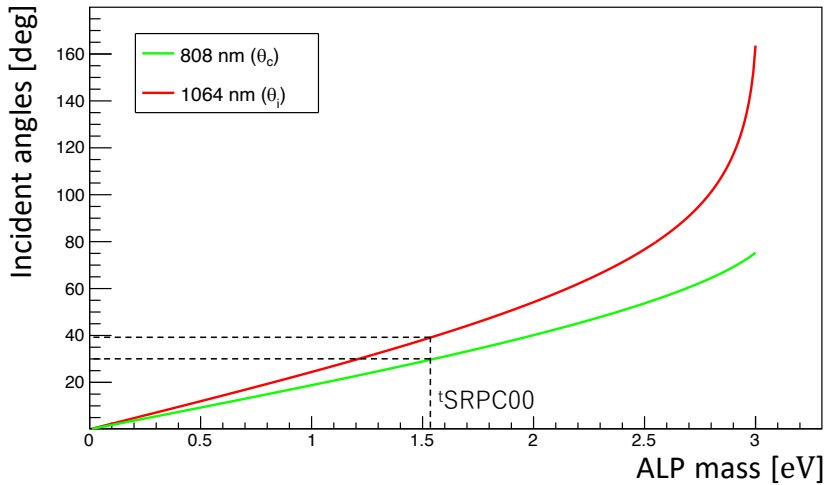}
\caption{
Expected incident angles of creation and inducing lasers, $\theta_c$ and $\theta_i$, respectively,
as a function of ALP mass when two wavelengths of 
creation (808 nm) and inducing lasers (1064 nm) are assumed 
resulting in the fixed wavelength of FWM signals, 651 nm, in vacuum.
}
\label{Fig11}
\end{figure}

The pilot search was indeed performed at a narrow mass range indicated by the angle points
as a function of ALP mass as shown in Fig.\ref{Fig11}.
Our prospect is to cover the broad mass range in the eV scale~\cite{3beam00}.
Toward the continuous mass scanning over the eV range with much higher laser intensity
in the near future,
the technical solutions developed in this pilot search will enable a realistic designing
for a more compact ${}^\mathrm{t}$SRPC operational in a vacuum chamber.

\section*{Acknowledgments}
The $T^{6}$ system was financially supported by
the MEXT Quantum Leap Flagship Program (JPMXS0118070187) and
the program for advanced research equipment platforms (JPMXS0450300521).
K. Homma acknowledges the support of the Collaborative Research
Program of the Institute for Chemical Research at Kyoto University
(Grant Nos.\ 2018--83, 2019--72, 2020--85, 2021--88, and 2022--101)
and Grants-in-Aid for Scientific Research
Nos.\ 17H02897, 18H04354, 19K21880, and 21H04474 from the Ministry of Education,
Culture, Sports, Science and Technology (MEXT) of Japan.
Y. Kirita acknowledges support from the JST, the establishment of university fellowships for the creation of science technology innovation, Grant No. JPMJFS2129, and a Grant-in-Aid for JSPS fellows No. 22J13756 from the Ministry of Education, Culture, Sports, Science and Technology (MEXT) of Japan.

\newpage

\section*{Appendix: Two angle parameters for Jones vectors representing general polarization states}

        \begin{figure}
        \centering
        \includegraphics[width=0.8\textwidth]{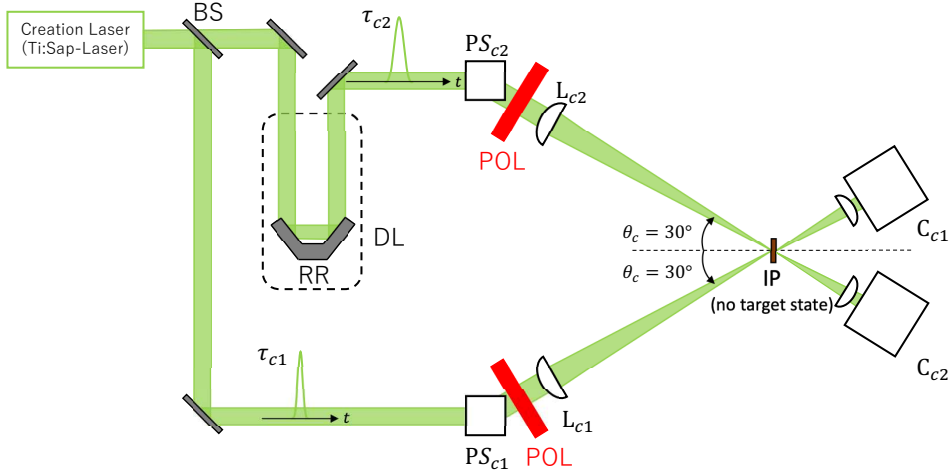}
        \caption{
Schematic view of the setup to evaluate two angle parameters to define Jones vectors
for the two creation lasers based on Stokes parameters.
A polarizer (POL) was placed between a periscope (PS) and a lens (L) in each beamline. 
Stokes parameters were obtained by measuring transmitted laser intensity through POL
at four different rotation angles by using individual cameras C$_{c1}$ and C$_{c2}$
assigned for the two creation lasers.
}
        \label{FigPol}
        \end{figure}

As mentioned in Section~\ref{Sec2}, incident angles of individual lasers were 
set at $30^\circ$ for the two creation lasers
with respect to the horizontal dashed line including IP as shown in Fig.\ref{FigPol} 
through periscopes PS$_{c1}$ and PS$_{c2}$, respectively.
We adopted PS to introduce a large collision angle, 
because it could reflect a laser beam to any angles 
changing the beam height thanks to a vertical pair of mirrors with
incident angle of $45^\circ$ and reflection angle of $45^\circ$ inside PS.
However, if the output direction is rotated by PS to a large angle, 
a linear polarization state of an incident laser beam becomes elliptically polarized.
Therefore, it is necessary to measure Stokes parameters to obtain 
ellipticity angle $\epsilon$ and tilt angle $\theta$ of a complex Jones vector
defined as follows that represents elliptically polarized states in general
        \begin{equation}
        \begin{pmatrix}
        \label{Jones}
        e_{1}\\
        e_{2}\\
        \end{pmatrix}
        =
        \begin{pmatrix}
        \cos\theta & -\sin\theta \\
        \sin\theta &  \cos\theta \\
        \end{pmatrix}  
        \begin{pmatrix}
        \cos\epsilon \\
        -i\sin\epsilon \\
        \end{pmatrix}  
        =~
        \begin{pmatrix}
        \cos\theta\cos\epsilon + i\sin\theta\sin\epsilon \\
        \sin\theta\cos\epsilon - i\cos\theta\sin\epsilon \\
        \end{pmatrix}
        .
        \end{equation}
Complex Jones vectors were actually implemented to polarization vectors
in four-vector form $e \equiv (0, e_1, e_2, 0)$
to define vertex factors for the ALP-photon coupling
in the numerical calculation to obtain volume-wise interaction rates
$\bar{\Sigma}_I$ in Eq.\eqref{yield} (see Ref.\cite{UNIVERSE} in more detail).

To obtain these two angles in complex Jones vectors for the two creation lasers, 
a polarizer (POL) was placed between a periscope (PS) and a lens (L) 
for each of the two creation laser lines as shown in Fig.\ref{FigPol}.
Rotation angles of POL around the optical axis were set to select linear polarization direction
of $0^{\circ}$(horizontal), $90^{\circ}$(vertical), $45^\circ$ and $135^\circ$. 
The Stokes parameters which can be converted into the two angle parameters for
complex Jones vectors were obtained by measuring laser intensities monitored 
by cameras (C$_{c1}$, C$_{c2}$) 
after laser lights pass through rotated POL set at the four rotation angles above.
A set of Stokes parameters can be related to two angle parameters of a complex Jones vector as follows
        \begin{equation}
        \begin{pmatrix}
        \label{stokes parameter_S}
        S_{0}\\
        S_{1}\\
        S_{2}\\
        S_{3}\\
        \end{pmatrix}
        =
        \begin{pmatrix}
        |E_{H}|^2 + |E_{V}|^2\\
        |E_{H}|^2 - |E_{V}|^2 \\
        |E_{45^\circ}|^2 - |E_{135^\circ}|^2\\
        |E_{L}|^2 - |E_{R}|^2 \\
        \end{pmatrix}  
        =~ S_0
        \begin{pmatrix}
        1\\
        \cos{2 \epsilon} \cos{2 \theta} \\
        \cos{2 \epsilon} \sin{2 \theta}\\
        \sin{2 \epsilon} \\
        \end{pmatrix}
        \end{equation}
where $E_{H}, E_{V}, E_{45^\circ}, E_{135^\circ}, E_{L}, E_{R}$ are 
the amplitudes for linear polarization cases with the polarization direction of horizontal, vertical, 
$45^\circ$, $135^\circ$ and for left- and right-handed circular polarization cases, respectively.
In the search experiment, we did not measure the right-handed and left-handed laser amplitudes, 
because we have only to obtain the two angle parameters: $\epsilon_k$ and $\theta_k$ for $k = c1, c2$
from $S_0, S_1$ and $S_2$.

The two angle parameters for the Jones vector in the creation laser path without retroreflector (RR) 
(lower side of the green optical path in Fig.\ref{FigPol}) were 
$\epsilon_{c1} = 0.41$ rad, $\theta_{c1} = 0.30$ rad, while
those in the path with RR were $\epsilon_{c2} = 0.91$ rad, $\theta_{c2} = - 0.31$ rad
as summarized in the following relation
        \begin{equation}
        \begin{pmatrix}
        \label{Aline_pol}
        S_{c10}\\
        S_{c11}\\
        S_{c12}\\
        S_{c13}\\
        \end{pmatrix}
        =~ A_{c1}^2
        \begin{pmatrix}
        1\\
        \cos{2 \epsilon_{c1}} \cos{2 \theta_{c1}} \\
        \cos{2 \epsilon_{c1}} \sin{2 \theta_{c1}}\\
        \sin{2 \epsilon_{c1}} \\
        \end{pmatrix}  
        =~ A_{c1}^2
        \begin{pmatrix}
        1\\
        \cos{\left(0.82\right)} \cos{\left(0.60\right)} \\
        \cos{\left(0.82\right)} \sin{\left(0.60\right)}\\
        \sin{\left(0.82\right)} \\
        \end{pmatrix}
        \end{equation}

        \begin{equation}
        \begin{pmatrix}
        \label{Bline_pol}
        S_{c20}\\
        S_{c21}\\
        S_{c22}\\
        S_{c23}\\
        \end{pmatrix}
        =~ A_{c2}^2
        \begin{pmatrix}
        1\\
        \cos{2 \epsilon_{c2}} \cos{2 \theta_{c2}} \\
        \cos{2 \epsilon_{c2}} \sin{2 \theta_{c2}}\\
        \sin{2 \epsilon_{c2}} \\
        \end{pmatrix}  
        =~ A_{c2}^2
        \begin{pmatrix}
        1\\
        \cos{\left(1.82\right)} \cos{\left(- 0.62\right)} \\
        \cos{\left(1.82\right)} \sin{\left(- 0.62\right)}\\
        \sin{\left(1.82\right)} \\
        \end{pmatrix}
        \end{equation}
where $A^2_{c1, c2}$ correspond to intensities measured by $C_{c1,c2}$, respectively.
The ellipticity angle for the creation laser containing RR was closer to $\pi/4$
than that of the other creation laser, 
because the incident angle of $45^\circ$ and the reflection angle of $45^\circ$ were 
not guaranteed within RR, while the tilt angles were opposite to each other as expected.
Therefore, the two creation lasers indeed had very
different angle parameters and these factors were taken into account for the numerical calculation
to set the exclusion region.


\end{document}